\documentclass[twocolumn]{jpsj3}
\usepackage{txfonts}
\usepackage{bm}
\usepackage{comment}

\title{A stochastic method to compute the $L^2$ localisation landscape}

\author{Masataka Kakoi\thanks{kakoi@presto.phys.sci.osaka-u.ac.jp}, and Keith Slevin\thanks{slevin.keith.sci@osaka-u.ac.jp}}
\inst{Department of Physics, Graduate School of Science, Osaka University, Toyonaka, Osaka 560-0043, Japan} 

\abst{The $L^2$ localisation landscape of L. Herviou and J. H. Bardarson is a generalisation of the localisation landscape of M. Filoche and S. Mayboroda. We propose a stochastic method to compute the $L^2$ localisation landscape that enables the calculation of landscapes using sparse matrix methods. We also propose an energy filtering of the $L^2$ landscape which can be used to focus on eigenstates with energies in any chosen range of the energy spectrum. We demonstrate the utility of these suggestions by applying the $L^2$ landscape to Anderson's model of localisation in one and two dimensions, and also to localisation in a model of the quantum Hall effect.}
\begin{document}
\maketitle

\section{Introduction}

To gain greater insight into the properties of the eigenstates of electrons in random potentials 
and the phenomenon of Anderson localisation,\cite{Anderson_1958}
Filoche and Mayboroda\cite{Filoche_2012} proposed solving a partial differential equation to find what they called a localisation landscape.
Suppose that the system is described by a Hamiltonian of the form
\begin{equation}
    H =  -\Delta + V(\bm{r}),
\end{equation}
where $V$ is typically a random potential. The eigenvalues $E_n$ and eigenstates $\psi_n$ are the solutions of
\begin{equation}\label{eq:eigenvalue_problem}
    H\psi_n(\bm{r}) = E_n\psi_n(\bm{r}).
\end{equation}
Instead of solving this eigenvalue problem, Filoche and Mayboroda suggest computing the localisation landscape which is the 
solution of
\begin{equation}\label{eq:original_landscape}
    H u(\bm{r}) = 1,
\end{equation}
with the same boundary conditions as Eq. (\ref{eq:eigenvalue_problem}).
Filoche and Mayboroda showed that, provided $V\geq 0$ everywhere (but not $V=0$ everywhere), the eigenvalues and eigenstates satisfy the inequality
\begin{equation}\label{eq:original_inequality}
    \left| \psi_n(\bm{r}) \right| \leq E_n u(\bm{r}).
\end{equation}
Note that for the condition on $V$ given above we have all $E_n>0$ and that here eigenstates are normalised so that the maximum value of the modulus is unity
\begin{equation}
    \|\psi_n\|_{\infty}=1.
\end{equation}
The landscape $u$ provides an upper bound for the modulus of eigenstates, particularly for lower energy eigenstates.
By looking at the landscape function we can discern where the eigenstates are localised.
In subsequent work it was shown that a localisation landscape can also be computed for discrete lattices.\cite{Lyra_2015, Wang_2021}
In this case for the inequality (\ref{eq:original_inequality}) to hold, all the elements of the inverse of the Hamiltonian matrix $H$ must be real and non-negative. 

Ways of overcoming the restrictions on $H$ have been proposed. For example, the construction of a landscape using 
comparison matrices.\cite{Lemut_2020}.
We focus here on the suggestion of Herviou and Bardarson\cite{Herviou_2020}.
In particular so that the localisation of excited states can be studied using a landscape,
Herviou and Bardarson defined what they called the $L^2$ landscape
\begin{equation}\label{eq:L2_landscape}
    u_i = \sqrt{(M^{-1})_{i,i}},
\end{equation}
where
\begin{equation}
    M = H^{\dagger}H.
\end{equation}
Here, $i$ labels sites on a lattice. For this landscape, with the aid of the Cauchy-Schwarz inequality, Herviou and Bardarson
showed that the eigenvalues and eigenstates of $H$ satisfy the following inequality
\begin{equation}\label{eq:L2_inequality}
    \left| \left( \psi_n \right)_i \right| \leq \left| E_n \right| u_i.
\end{equation}
Unlike in the approach of Filoche and Mayboroda, here the wave function is normalised so that the $L^2$ norm is unity
\begin{equation}\label{eq:2_norm}
    \|\psi_n\|_{2}=1.
\end{equation}
The only restriction on the Hamiltonian $H$ is now the existence of an inverse for $M$. 

In terms of the eigenvalues and eigenstates of $H$ the square of the $L^2$ landscape can be expressed as
\begin{equation}\label{eq:L2_expansion}
    u_i^2 = \sum_n \frac{1}{\left| E_n \right|^2 }|(\psi_n)_i|^2.
\end{equation}
This expression makes clear that states near zero energy have a heavier weight in the landscape.
To study states near a particularly energy Herviou and Bardarson introduced an energy shift $E_{\mathrm{shift}}$, i.e., they replace
$H$ in Eq. (\ref{eq:L2_landscape}) with
\begin{equation} \label{eq:def_energy_shift}
    h = H - E_{\mathrm{shift}}I,
\end{equation}
where $I$ is the identity matrix.
Moreover, since the inequality Eq. (\ref{eq:L2_inequality}) holds even if $H$ is not Hermitian, Herviou and Bardarson also suggested introducing a small imaginary part in the energy shift. This ensures that $M$ is always invertible.

In this paper, we propose a stochastic method to compute the $L^2$ landscape that allows efficient sparse matrix methods to be used.
We also introduce a Butterworth-type energy filter\cite{Butterworth_1930} to make it easier to focus on states in arbitrary parts of the spectrum of the Hamiltonian.
We apply these methods to Anderson's model of localisation in one and two dimensions. We also simulate Anderson's model of localisation in two dimensions subject to a strong uniform perpendicular magnetic field that puts the system in the regime of the quantum Hall effect. In this later case we apply the method to localisation in the lowest Landau band.

\section{Theory and Method}

\subsection{Stochastic method for the $L^2$ localisation landscape}

We replace the right-hand side of Eq. (\ref{eq:original_landscape}) with a random vector $\bm{\xi}$
\begin{align}
    \label{eq:L2_stochastic}
    h\bm{w} = \bm{\xi}.
\end{align}
Here, the distribution of the elements of the vector $\xi_i$ should satisfy the following conditions
\begin{align}
    \label{eq:random_vectors_mean}
    \left<\xi_i\right> &= 0,\\
    \label{eq:random_vectors_var}
    \left<\xi_i\xi_j^*\right> &= \delta_{i,j},
\end{align}
Inverting Eq. (\ref{eq:L2_stochastic}) we have
\begin{equation}
    w_i = \sum_{j} \left(h^{-1}\right)_{i,j} \xi_j,
\end{equation}
so that
\begin{equation}
     |w_i|^2 = \sum_{j,k} (h^{-1})_{i,j} \xi_j (h^{-1})_{i,k}^* \xi_k^*.
\end{equation}
Now, using Eqs. (\ref{eq:random_vectors_mean}) and (\ref{eq:random_vectors_var}) to calculate the expectation value of the square of the modulus of the solution $w_i$, we find that it is equal to the square of the $L^2$ localisation landscape Eq. (\ref{eq:L2_landscape})
\begin{align}\label{eq:expectation_value}
    \nonumber
    \left\langle |w_i|^2\right\rangle &= \sum_{j}(h^{-1})_{i,j}(h^{-1})_{i,j}^*\\
    \nonumber
    &= \sum_{j}(h^{-1})_{i,j} ((h^{-1})^{\dag})_{j,i}\\
    \nonumber
    &= (M^{-1})_{i,i}\\
    &= u_i^2.
\end{align}

We use two distributions of random vectors satisfying Eqs. (\ref{eq:random_vectors_mean}) and (\ref{eq:random_vectors_var}).
For the first distribution the elements $\xi_i$ of the random vectors are independently
and identically distributed with a normal distribution with mean zero and variance one.
Since these are real vectors, this is an appropriate choice, if $h$ is a real matrix.

For the second distribution the elements $\xi_i$ of the random vectors have absolute value unity and
random phase $\theta_i$ independently and identically distributed with a uniform distribution 
on the interval $\left[0,2\pi\right]$, i.e.,
\begin{equation} \label{eq:random_phase}
    \xi_i = \exp\left(\mathrm{i} \theta_i \right).
\end{equation}
When $h$ is a complex matrix we use random vectors distributed in this way.

We approximate the expectation value Eq. (\ref{eq:expectation_value}) using the average over $N$ independently sampled vectors $\bm{\xi}^{(\alpha)}$, i.e.,
\begin{equation}\label{eq:linear_equation}
     h\bm{w}^{(\alpha)} = \bm{\xi}^{(\alpha)},
\end{equation}
where $\alpha =1, \ldots , N$ and then
\begin{equation}
    u_i^2 \approx \overline{|w_i|^2} = \frac{1}{N}\sum_{\alpha=1}^{N}\left|w_i^{(\alpha)}\right|^2.
\end{equation}
We choose the number of samples $N$  so as to obtain the landscape within a desired precision.
The standard deviation of the statistical error in this estimate decreases as $1/\sqrt{N}$. 
Note that the statistical fluctuations in the estimate of the
landscape at different sites are correlated. The calculation of the full co-variance matrix is given in Appendix\ref{ErrorCalculation}.

We have implemented the calculations using the Python programming language.
When $h$ is Hermitian and sparse, a method such as the conjugate gradient method is appropriate when solving Eq. (\ref{eq:linear_equation}).
We first precondition the matrix using incomplete LU factorisation (ILU) 
with the scipy.sparse.linalg.ilu routine and then solve the linear equations using scipy.sparse.linalg.cg
of the SciPy library.

\subsection{Energy filtering}

\begin{figure}[tbp]
  \centering
  \includegraphics[width=8cm]{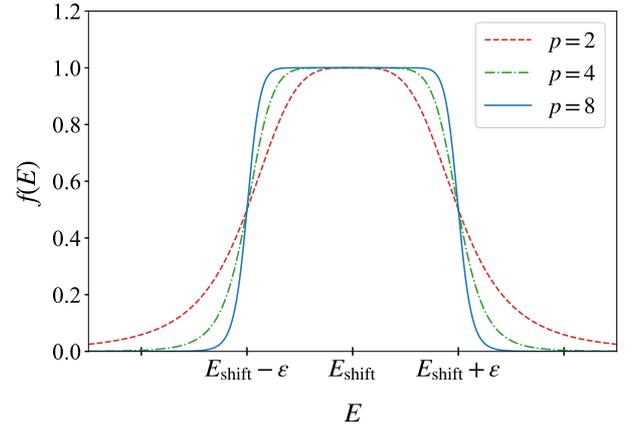}
  \caption{(Colour online) Form of the Butterworth-type filter $f(E)$ in Eq. (\ref{eq:filter_func}). For sufficiently large $p$, the function approaches $0$ for $|E|>\varepsilon$ and $1$ for $|E|<\varepsilon$.}
  \label{fig:filter_func}
\end{figure}

To focus more precisely on states near a chosen energy we propose the use of a Butterworth-type filter\cite{Butterworth_1930}
when calculating the landscape.
In particular, consider the function
\begin{equation}
    \label{eq:filter_func}
     f(E) = \left[ \left(\frac{E-E_{\mathrm{shift}}}{\varepsilon}\right)^{2p} + 1 \right]^{-1}.
\end{equation}
Here $\varepsilon>0$ and $p$ is a positive integer. This function is plotted in Fig. \ref{fig:filter_func} for various values of $p$.
In place of Eq. (\ref{eq:def_energy_shift}) we set
\begin{equation}
    h = f^{-1}(H) = h_0^{2p} + I,
\end{equation}
where
\begin{equation}
    h_0 = \frac{H-E_{\mathrm{shift}}I}{\varepsilon}.
\end{equation}
This matrix $h$ has the same eigenstates as $H$ and has eigenvalues $1/f(E_{n})$.
The theory of Herviou and Bardarson is equally applicable to $h$ as $H$, so we arrive at the following inequality for the eigenstates and eigenvalues of $H$
\begin{equation} \label{eq:L2_filtered_inequality}
f(E_n)\left|(\psi_n)_i\right| \leq u_i,
\end{equation}
and the following expression for the filtered landscape in terms of the eigenvalues and eigenstates of $H$
\begin{equation} \label{eq:L2_filtered_expansion}
    u_i^2 = \sum_n \left(f(E_n)\right)^2|(\psi_n)_i|^2.
\end{equation}
From this equation, and referring to Fig. \ref{fig:filter_func}, we see that the contribution to the landscape of eigenstates with energies outside the filter window, which has width approximately $2\epsilon$ and is centred at energy $E_{\mathrm{shift}}$, are strongly suppressed.
From Eq. (\ref{eq:L2_filtered_inequality}), we see that the filter provides constraints on the eigenstates inside the filter window, and no constraints on the eigenstates outside the window.
The sharpness of the filter is set by the parameter $p$.
In what follows, we set $p=8$.

\subsection{Avoiding underflow or overflow}

We solve Eq. (\ref{eq:L2_stochastic}) in steps to avoid underflow or overflow.
To do so we make use of the identity
\begin{equation}
    q^{2p}+1 = \prod_{k=1}^{2p} \left( q + q_k \right),
\end{equation}
where $q_1,\cdots ,q_{2p}$ are the $2p$th roots of $-1$, i.e.,
\begin{equation}\label{eq:qk}
    q_k = \exp\left(\mathrm{i}\frac{2k-1}{2p}\pi\right).
\end{equation}
Setting
\begin{equation}
    h_k =  h_0 + q_k I,
\end{equation}
we then have
\begin{equation}
    \label{eq:factorization_complex}
    h_0^{2p} + I= \prod_{k=1}^{2p} h_k.
\end{equation}
We can now solve Eq. (\ref{eq:L2_stochastic}) in $2p$ steps as follows
\begin{align}
    \bm{v}_0 &= \bm{\xi},\\
    \label{eq:numerical_stochastic_filter_2}
    h_k\bm{v}_k &= \bm{v}_{k-1}\ \ \ \ \ \ \mathrm{for}\ k=1,2,\ldots,2p,\\
    \bm{w} &= \bm{v}_{2p},
\end{align}
Unfortunately, this procedure increases the computational load in two ways. 
First, for each vector ${\boldsymbol{\xi}}$ we need to solve $2p$ sets of linear
equations. 
Second, since the $h_k$  in Eq. (\ref{eq:numerical_stochastic_filter_2}) are not Hermitian, 
the conjugate gradient method is not appropriate and 
we use the stabilised bi-conjugate gradient method instead
as implemented in routine scipy.sparse.linalg.bicgstab.
This increase in computational load can be mitigated to a certain extent by modifying the algorithm 
to render the matrices Hermitian. This is described in Appendix \ref{alternative_method}.

\section{Results}

\subsection{Anderson’s model of localisation in one dimension}

We apply the stochastic method first to Anderson’s model of localisation\cite{Anderson_1958} in one dimension,
\begin{equation} \label{eq:AndersonModel}
    H = \sum_i W_i c^{\dag}_i c^{}_i -t\sum_{\langle i,j\rangle}c^{\dag}_i c^{}_j.
\end{equation}
Here, $i$ labels sites with position $x_i$ on a one-dimensional lattice with lattice constant $a$, 
$c^{\dag}_i$ ($c^{}_i$) is the creation (annihilation) operator on site $i$, and the
$W_i$ are independently and identically distributed random variables with uniform distribution on the interval $[-W/2, W/2]$. 
The second term is summed over pairs of nearest-neighbour sites. 
We take the distance $a$ between sites as a unit of length and the hopping $t$ as a unit of energy. 
In this paper we always use periodic boundary conditions.

\begin{figure*}[t!]
    \centering
    \includegraphics[width=14.5cm]{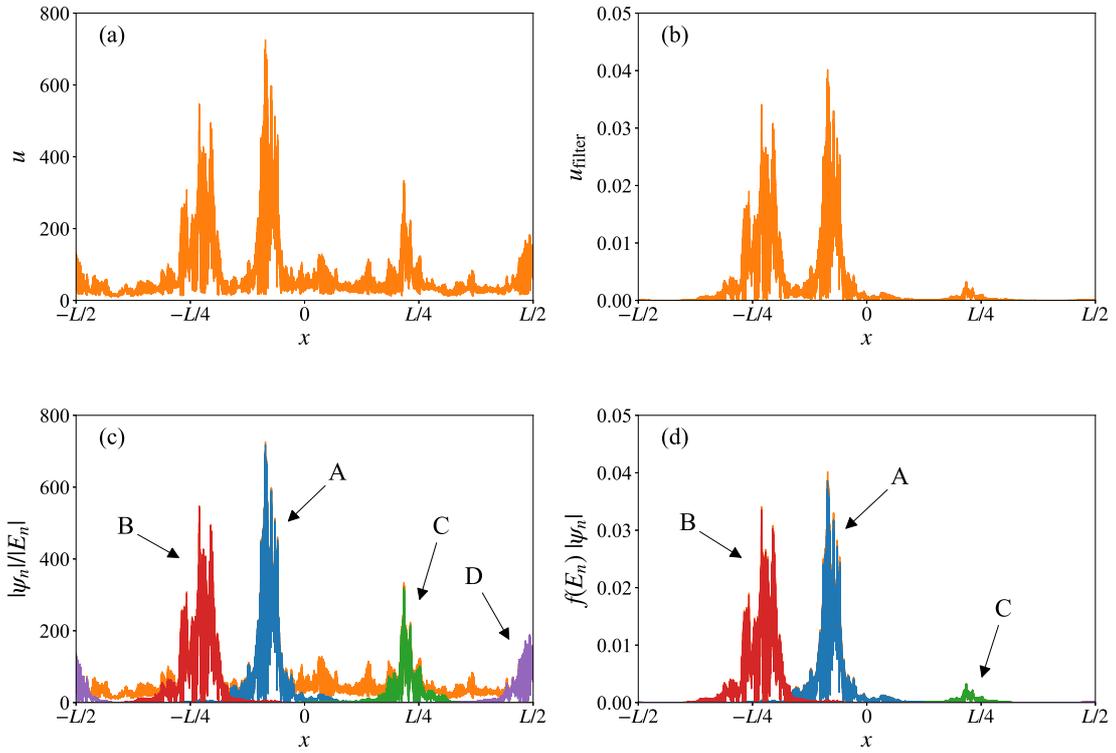}
    \caption{(Colour online) Comparison of eigenstates and landscapes with $E_{\mathrm{shift}}=0$, i.e., near the centre of the band. (a) $L^2$ landscape without filtering obtained by the stochastic method. (b) $L^2$ landscape with filtering obtained by the stochastic method. The filter parameters are $\varepsilon=1.2\times10^{-4}$ and $p=8$. (c) and (d) The four eigenstates (calculated using the Lanczos algorithm) with energies closest to $E_{\mathrm{shift}}$ are overlaid on the landscape from figures (a) and (b), respectively. 
    The eigenstates are labelled A, B, C, and D in order of energy closest to $E_{\mathrm{shift}}$. The energies are $-5\times10^{-5}$, $6\times10^{-5}$, $-1.4\times10^{-4}$, and $1.7\times10^{-4}$, respectively.}
    \label{fig:landscape_1D}
\end{figure*}

Figure \ref{fig:landscape_1D} shows the results of calculating the $L^2$ localisation landscape for a system with $L=2^{16}=65,536$ sites and disorder strength $W=0.2$. 
In Fig. \ref{fig:landscape_1D}(a), we show the $L^2$ landscape computed using the stochastic method with $E_{\mathrm{shift}}=0$ and without filtering.
We used random vectors with normally distributed elements and the sampling of random vectors was stopped when the 
maximum statistical error reached 5\%. This required $N=1160$ vectors.
From Eq. (\ref{eq:L2_filtered_expansion}), we expect the largest contribution to the unfiltered $L^2$ landscape
to come from eigenstates near $E=E_{\mathrm{shift}}$, i.e., $E=0$.
We also expect that the inequality Eq. (\ref{eq:L2_inequality}) to be closest to being saturated for eigenstates at the same energy.
This is consistent with what we see in Fig. \ref{fig:landscape_1D}(c) where the four eigenstates with energies closest to $E=0$ are superimposed on the landscape.

In Fig. \ref{fig:landscape_1D}(b), we show a filtered $L^2$ landscape with $\varepsilon=1.2\times10^{-4}$ and $p=8$.
In a separate calculation using the Kernel Polynomial Method\cite{Weisse_2006} (KPM) we estimated the density of states (DOS) at $E=0$ to
be $\rho\approx 0.159$ per site. Thus, the value of $\varepsilon$ used in the filter corresponds to approximately 1.25 times the
mean energy level spacing.
We used random vectors with elements with uniformly distributed random phases and the sampling of random vectors was stopped when the 
maximum statistical error reached 5\%. This required $N=470$ vectors.
In Fig. \ref{fig:landscape_1D}(d), we show the three eigenstates with energies inside the filter window superimposed on the 
filtered $L^2$ landscape shown in Fig. (2b).
We see that the contribution to the landscape from eigenstates outside the filter window has been successfully filtered out
and that the landscape inequality for the filtered landscape Eq. (\ref{eq:L2_filtered_inequality}) is almost saturated.

\subsection{Anderson’s model of localisation in two dimensions}

We now turn to Anderson's model of localisation in two dimensions. The Hamiltonian is formally the same as 
Eq. (\ref{eq:AndersonModel}) but the index $i$ now labels sites with position $\left(x_i, y_i\right)$ 
on a two-dimensional square lattice with lattice constant $a$. 
We simulated an $L \times L$ system with $L=2^{8}=256$ (in units of $a$) and disorder $W=6$ (in units of $t$). 

\begin{figure*}[t!]
    \centering
    \includegraphics[width=11.5cm]{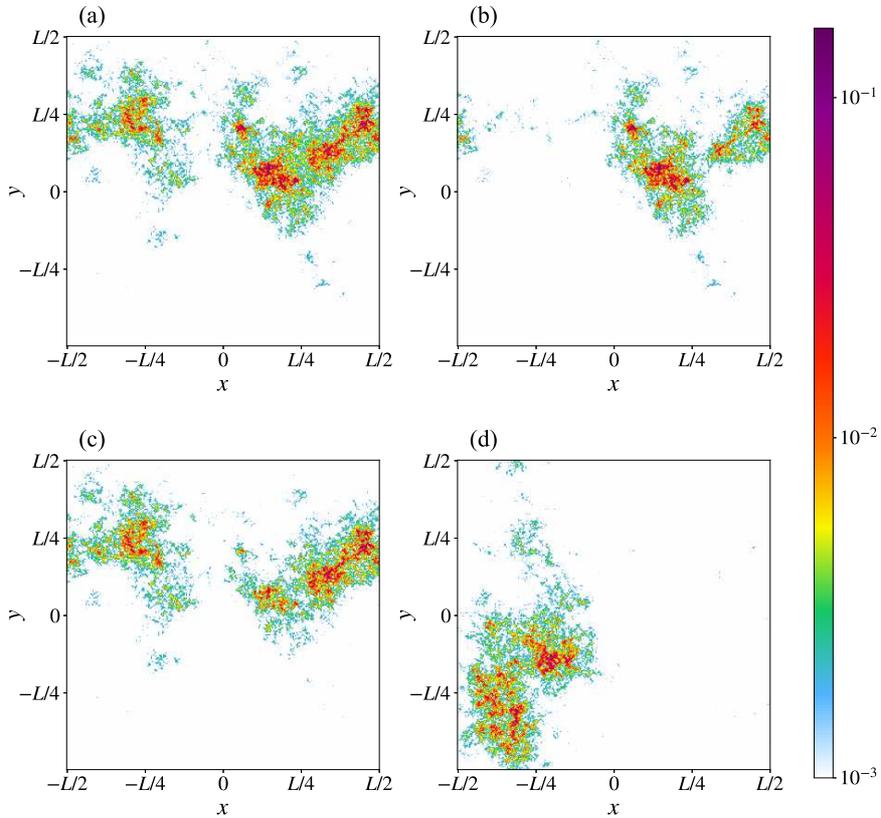}
    \caption{(Colour online) $L^2$ landscape and three eigenstates for Anderson’s model of localisation in two dimensions with linear system size $L=256$ and disorder $W=6$: (a) $L^2$ filtered landscape with $E_{\mathrm{shift}}=0$, $\varepsilon=1.25\times10^{-4}$, and $p=8$. (b) Modulus of an eigenstate near the band centre with energy $E=9.7\times10^{-5}$. (c) An eigenstate with energy $E=-1.12\times10^{-4}$. (d) An eigenstate with energy $E=-2.6\times10^{-4}$, an energy that is outside the filter window.}
    \label{fig:landscape_2D}
\end{figure*}

In Fig. \ref{fig:landscape_2D}(a), we show the filtered $L^2$ landscape computed using the stochastic method with 
$E_{\mathrm{shift}}=0$, $\varepsilon=1.25\times10^{-4}$, and $p=8$.
We used random vectors with elements with uniformly distributed random phases and the sampling of random vectors was stopped when the 
maximum statistical error reached 5\%. This required $N=490$ vectors.
Using the KPM method we estimated the DOS at $E=0$ to be $\rho\approx 0.123$ per site.
Thus, the value of $\varepsilon$ used in the filter corresponds to approximately the mean energy level spacing.
Hence, we expect to find on average approximately two eigenstates within the filter window.
In Fig. \ref{fig:landscape_2D}(b)-(d), we show eigenstates with energies $E=9.7\times10^{-5},\ -1.12\times10^{-4}$, and $-2.6\times10^{-4}$, respectively.
Comparison of the landscape and eigenstates shows that the amplitude of the $L^2$ landscape accurately indicates the 
spatial location of the eigenstates that have energies within the filter window and also that eigenstates with energies
outside the filter window are successfully filtered out.
Estimates of $37a$ and $48a$ for the localisation length at the band centre 
for this model and parameters have been reported in Refs. \citenum{MacKinnon_1983} and \citenum{Slevin_2004}, respectively.
The results shown are consistent with these estimates. 

\subsection{Two dimensional system in a uniform perpendicular magnetic field}

\begin{figure}[tbp]
  \centering
  \includegraphics[width=8.5cm]{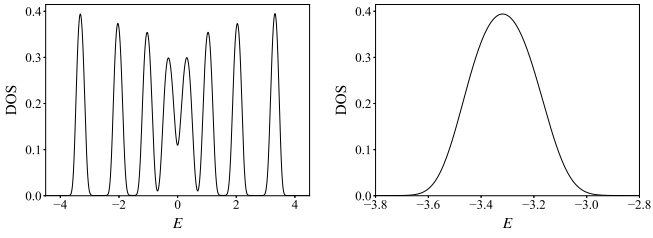}
  \caption{The DOS per site of the model defined in Eq. (\ref{eq:magnetic_field_model}) with parameters $\phi=1/8$ and $W=1$. (left) Full DOS. (right) Enlarged view of the DOS near the lowest Landau sub-band.}
  \label{fig:magnetic_DOS}
\end{figure}

We now apply the $L^2$ landscape to Anderson's model of localisation in two dimensions subject to a uniform perpendicular magnetic field.
The Hamiltonian is
\begin{equation} \label{eq:magnetic_field_model}
    H = \sum_i W_i c^{\dag}_i c^{}_i -t\sum_{\langle i,j\rangle} \exp\left(\mathrm{i} \varphi_{ij} \right) c^{\dag}_i c^{}_j.
\end{equation}
The index $i$ labels sites with position $\left(x_i, y_i\right)$ on a two-dimensional square lattice with lattice constant $a$. 
We suppose that a uniform magnetic field is applied perpendicular to the system.
The vector potential $\boldsymbol{A}=\left(0, Bx, 0\right)$ corresponds to a magnetic flux $B$ in the $+z$ direction.
The application of a magnetic field in a lattice model is described by introducing phases $\varphi_{ij}$ (Peierls factors\cite{Peierls_1933})
in the hopping terms. These phases are
\begin{equation}
    \varphi_{ij} = 0,
\end{equation}
when site $j$ is a nearest neighbour of site $i$ in the $\pm x$ directions and
\begin{equation}
    \varphi_{ij} = \pm 2\pi \phi x_i,
\end{equation}
when site $j$ is a nearest neighbour of site $i$ in the $\pm y$ directions.
Here, $\phi$ is the magnetic flux threading a 
unit cell of the square lattice in units of the flux quantum $h/e$, i.e.,
\begin{equation}
    \phi = \frac{Ba^2}{h/e}.
\end{equation}
When $\phi$ is a rational number, i.e., $\phi=p/q$, $q$ Landau bands appear\cite{Hofstadter_1976}.
These have an intrinsic broadening.
In the presence of disorder $W>0$ the Landau bands are further broadened.
Also states in the Landau bands are Anderson localised except for a single critical energy $E_c$ close to the centre of the sub-band.
Provided that the disorder $W$ is not too large the Landau bands remain resolved and the system is then in the regime of the quantum Hall effect (QHE).
We have simulated a square system with $L \times L$ sites with linear size $L=256$, disorder $W=1$, and
magnetic flux $\phi = 1/8$ flux quantum per unit cell.
To confirm that the system is in the QHE regime we show in Fig. \ref{fig:magnetic_DOS} the DOS calculated using the KPM\cite{Weisse_2006}. 
In what follows we focus on the sub-band with lowest energy.

\begin{figure*}[t!]
    \centering
    \includegraphics[width=11.5cm]{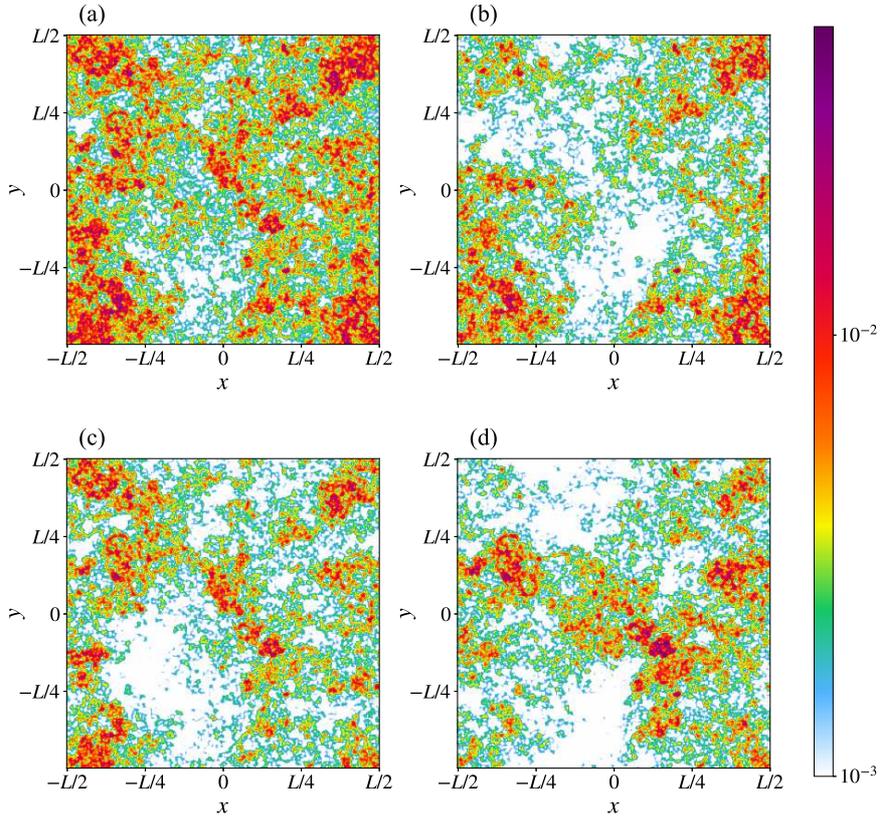}
    \caption{(Colour online) $L^2$ landscape and three eigenstates for a square system in a perpendicular magnetic field.
    The system has $L \times L$ sites with linear size $L=256$, disorder $W=1$, and magnetic flux $\phi = 1/8$ flux quantum per unit cell: 
    (a) Filtered $L^2$ landscape with $E_{\mathrm{shift}}=-3.3$, $\varepsilon=4\times10^{-5}$, and $p=8$.
    This value of $E_{\mathrm{shift}}$ is at the centre of the first Landau band.
    (b)-(d) Modulus of eigenstates with energies $E=E_{\mathrm{shift}}+1.4\times10^{-5}$, $E=E_{\mathrm{shift}}-2.2\times10^{-5}$, and $E=E_{\mathrm{shift}}+5.1\times10^{-5}$, respectively.}
    \label{fig:magnetic_landscape_A}
\end{figure*}

We report two calculations of the landscape with a magnetic field.
In the first we set $E_{\mathrm{shift}}=-3.3$, close to the centre of the Landau band,
and $\varepsilon=4\times10^{-5}\approx$ mean level spacing.
The filter window is expected to contain approximately two eigenstates.
We used random vectors with elements with uniformly distributed random phases and the sampling of random vectors was stopped when the 
maximum statistical error reached 5\%. This required $N=510$ vectors.
In Fig. \ref{fig:magnetic_landscape_A}(b)-(d), we show eigenstates with energies 
$E=E_{\mathrm{shift}}+1.4\times10^{-5}$, $E= E_{\mathrm{shift}}-2.2\times10^{-5}$, and $E=E_{\mathrm{shift}}+5.1\times10^{-5}$, respectively. 
The eigenstates in (b) and (c) have energies inside the filter window, while the eigenstate in (d) 
has an energy outside the window and does not contribute to the landscape.
As is well known, the localisation length diverges at the centre of the Landau bands, and this is
clearly reflected in the landscape. 

\begin{figure*}[t!]
    \centering
    \includegraphics[width=11.5cm]{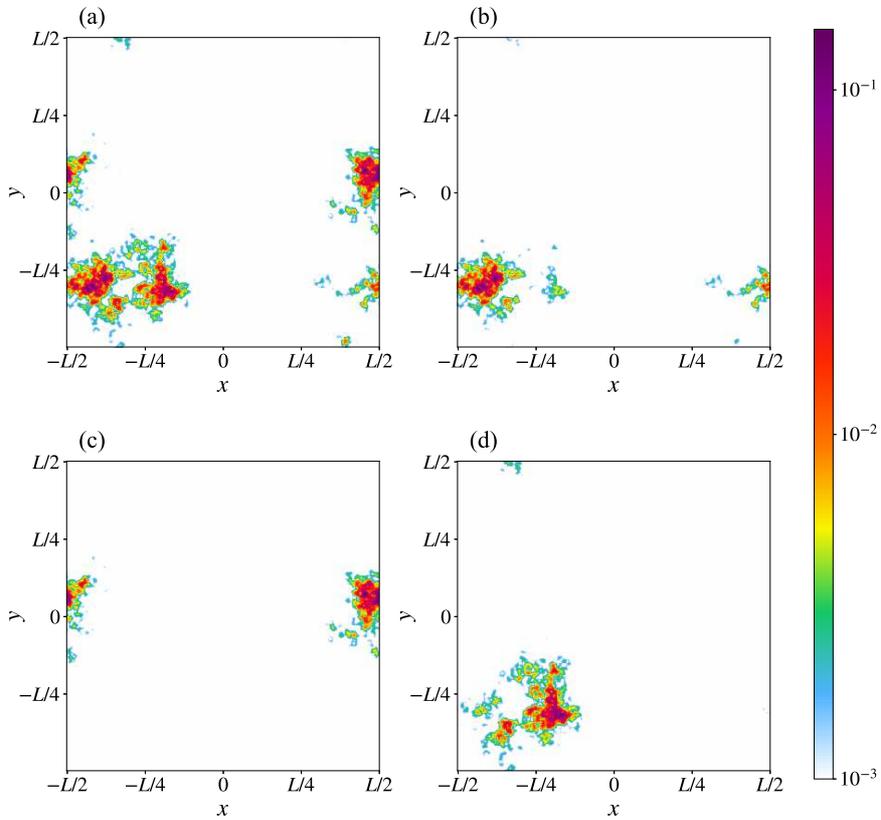}
    \caption{(Colour online) 
    As for Figure \ref{fig:magnetic_landscape_A} but with $E_{\mathrm{shift}}=-3.2$, $\varepsilon=10^{-5}$, and $p=8$.
    (a) Filtered $L^2$ landscape,
    (b)-(d) Modulus of eigenstates with energies $E=E_{\mathrm{shift}}-1.4\times10^{-5}$, $E=E_{\mathrm{shift}}-2.8\times10^{-5}$, and $E=E_{\mathrm{shift}}-4.7\times10^{-5}$, respectively.}
    \label{fig:magnetic_landscape_B}
\end{figure*}

In the second calculation we set $E_{\mathrm{shift}}=-3.2$, away from the centre of the Landau band, and
$\varepsilon=5\times10^{-5}\approx$ mean level spacing.
We used random phase distributed random vectors and the sampling of random vectors was stopped when the 
maximum statistical error reached 5\%. 
This required $N=520$ vectors.
In Fig. \ref{fig:magnetic_landscape_B}(b)-(d), we show eigenstates with energies $E=E_{\mathrm{shift}}+1.4\times10^{-5}$, $E=E_{\mathrm{shift}}-2.8\times10^{-5}$, and $E=E_{\mathrm{shift}}-4.7\times10^{-5}$, respectively.
Since all these energies are inside the filter window, the peaks in the landscape correspond with the three eigenstates.

\subsection{Comparison between the stochastic method and the Lanczos algorithm}

Since the stochastic method involves repeatedly solving linear equations, its computational complexity is similar to that of a sparse matrix eigenvalue solver like the Lanczos algorithm\cite{Lanczos_1950}. 
In fact, to obtain a small number of eigenstates, the Lanczos algorithm is faster and provides a value for the energy of
each eigenstate. 
The stochastic method becomes more efficient when there are many eigenstates in the filter. 
In the stochastic method, increasing the width of the filter window does not significantly increase the computation time. 

We compare the stochastic method with the Lanczos algorithm for the centre of the band of Anderson's model of localisation in one-dimension with $L=2^{18}=262144$ sites and $W=2$.
We use $E_{\mathrm{shift}}=0$, $\varepsilon=0.02$, and $p=8$ as parameters for the stochastic method.
We estimate the DOS at $E=0$ to be $\rho\approx 0.161$ per site.
Therefore, we expect approximately $2\varepsilon \rho L\approx 1680$ eigenstates inside the filter.
We carried out calculations for 10 different realisations of the random potential.
We solved the eigenvalue problem using the scipy.sparse.linalg.eigsh function from the SciPy library, which is a wrapper for the ARPACK\cite{ARPACK_user_guide} SSEUPD and DSEUPD functions.
The CPU used was an Intel Xeon Silver 4208 with a processor base frequency of 2.10 GHz.
The average time required for the 10 calculations was 315 seconds and 2076 seconds for the stochastic method and Lanczos algorithm, respectively, with standard deviations of 12 seconds and 47 seconds, respectively.

\begin{figure}[tbp]
    \centering
    \includegraphics[width=8cm]{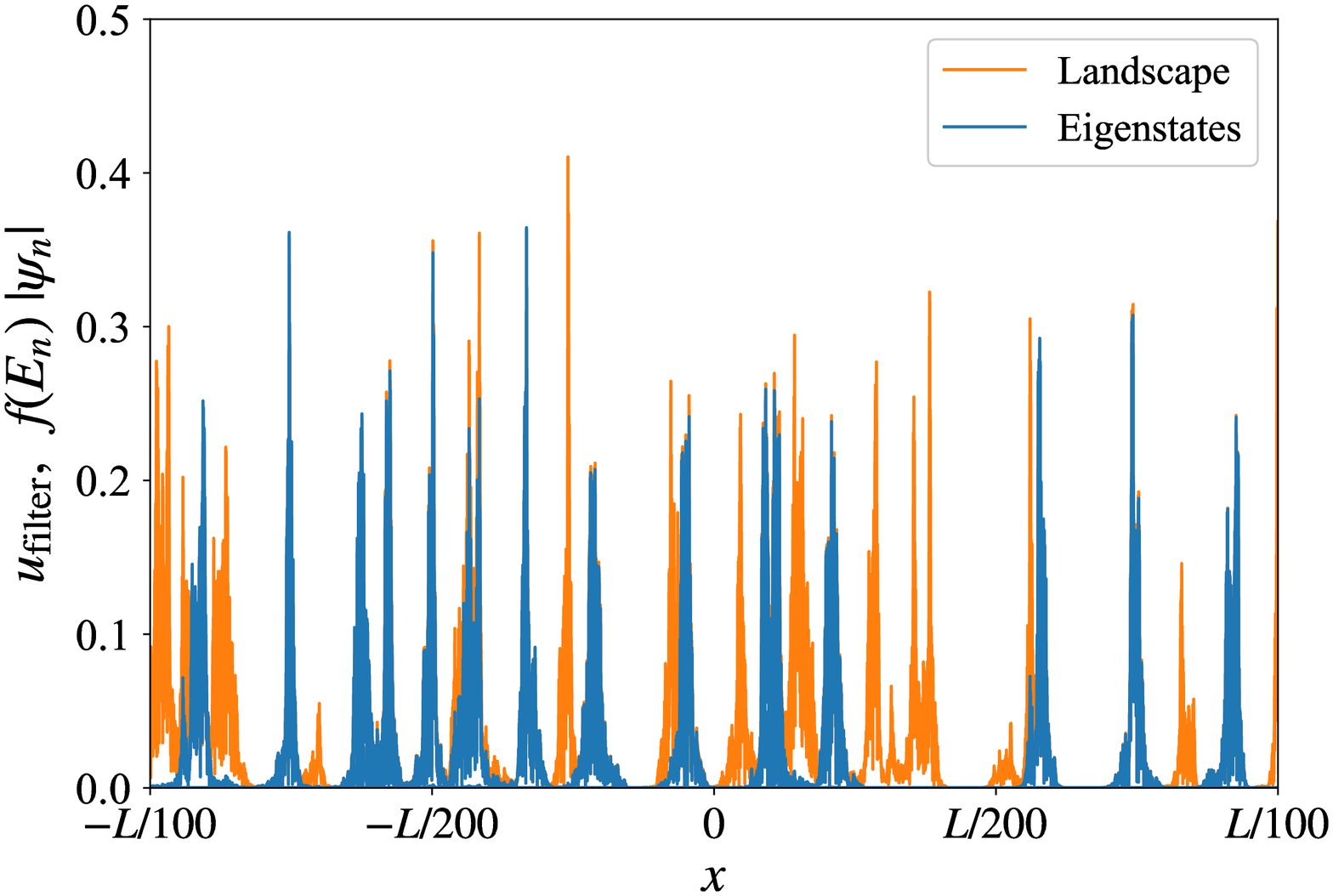}
    \caption{(Colour online) Comparison of eigenstates and landscape in an
enlarged view near the centre of the system.
The filter parameters are  $E_{\mathrm{shift}} = 0$, $\varepsilon=0.02$, and $p=8$. }
    \label{fig:timing}
\end{figure}

In Fig. \ref{fig:timing}, we compare eigenstates calculated using the Lanczos algorithm and the landscape in an
enlarged view near the centre of the system. 
Only 800 eigenstates have been calculated with the Lanczos algorithm, compared with the 1680 expected inside the filter,
so the landscape exhibits approximately twice as many peaks.
Thus, although the stochastic method does not provide as much information about each eigenstate as the Lanczos algorithm, it allows us to extract information about a larger number of eigenstates in much less time.

\section{Discussion}

We have suggested a stochastic method for estimating the $L^2$ landscape.
This method avoids a direct computation of the inverse of the Hamiltonian matrix. 
The stochastic method permits the application of sparse matrix methods and is ideally suited to parallel 
calculations since Eq. (\ref{eq:L2_stochastic}) can be solved in parallel for statistically independent streams of random vectors.
The way in which random vectors are employed here is very similar to their use in
the KPM\cite{Weisse_2006}.

We have also introduced an extension of the $L^2$ landscape with energy filtering.
This method filters out the contribution of eigenstates with energies outside the filter window and
permits a focus on eigenstates with energies near an arbitrary energy of interest.
This makes the landscape useful for the study of localisation not only near the edges of the band where the DOS is small but
to the centre of the band also where the DOS is large.

From the original landscape of Filoche and Mayboroda\cite{Filoche_2012}, an effective potential has been defined\cite{Arnold_2016}.
The integrated DOS for semiconductors has been investigated using this effective potential\cite{Filoche_2017,Piccardo_2017,Li_2017}.
The Wigner function has also been investigated \cite{Pelletier_2022,Banon_2022}.
It is not yet clear if an appropriate effective potential can be defined for the $L^2$ landscape and this may be an
interesting question for future research.

Another direction is the application of localisation landscapes to many-body problems.
For the original landscape, such an application has been proposed\cite{Balasubramanian_2020,Stellin_2023}.
Unlike the original landscape the $L^2$ landscape is applicable to wider range of Hamiltonians.
This, combined with the energy filtering we suggest here, may be advantageous in the study of localisation in a
many-body context.

\section*{Acknowledgment}

Keith Slevin is grateful for support from the Japan Society for the Promotion of Science under Grant-in-Aid 19H00658.


\bibliographystyle{jpsj}
\bibliography{references}

\onecolumn

\appendix

\section{} \label{ErrorCalculation}

To determine the  precision of the estimation of the $L^2$ landscape using the stochastic method we
calculate the covariance matrix of the landscape estimate, i.e., we calculate
\begin{align}
    \label{eq:u_i^2u_j^2}
    \nonumber
    \left\langle \overline{|w_i|^2}\cdot\overline{|w_j|^2} \right\rangle
    &= \frac{1}{N^2} \sum_{\alpha,\beta}\sum_{k,l,m,n} (h^{-1})_{i,k}^* (h^{-1})_{i,l} (h^{-1})_{j,m}^* (h^{-1})_{j,n} \left\langle \xi_k^{(\alpha)*} \xi_l^{(\alpha)} \xi_m^{(\beta)*} \xi_n^{(\beta)} \right\rangle \\
    \nonumber
    &= \frac{1}{N^2} \left[\sum_{\alpha\neq\beta} \sum_{k,l,m,n} (h^{-1})_{i,k}^* (h^{-1})_{i,l} (h^{-1})_{j,m}^* (h^{-1})_{j,n} \left\langle \xi_k^{(\alpha)*} \xi_l^{(\alpha)} \right\rangle \left\langle \xi_m^{(\beta)*} \xi_n^{(\beta)} \right\rangle\right.\\
    \nonumber
    & \quad \left.+ \sum_{\alpha} \sum_{k,l,m,n} (h^{-1})_{i,k}^* (h^{-1})_{i,l} (h^{-1})_{j,m}^* (h^{-1})_{j,n} \left\langle \xi_k^{(\alpha)*} \xi_l^{(\alpha)} \xi_m^{(\alpha)*} \xi_n^{(\alpha)} \right\rangle\right] \\
    \nonumber
    &= \left.\frac{N-1}{N} \sum_{k,m} (h^{-1})_{i,k}^* (h^{-1})_{i,k} (h^{-1})_{j,m}^* (h^{-1})_{j,m} \right.\\
    \nonumber
    & \quad + \frac{1}{N} \sum_{k,m} (h^{-1})_{i,k}^* (h^{-1})_{i,k} (h^{-1})_{j,m}^* (h^{-1})_{j,m} (1-\delta_{k,m}) \\
    \nonumber
    & \quad + \frac{1}{N} \sum_{k,l} (h^{-1})_{i,k}^* (h^{-1})_{i,l} (h^{-1})_{j,l}^* (h^{-1})_{j,k} (1-\delta_{k,l}) \\
    \nonumber
    & \quad + \frac{\left|\left\langle\xi^2\right\rangle\right|^2}{N} \sum_{k,l} (h^{-1})_{i,k}^* (h^{-1})_{i,l} (h^{-1})_{j,k}^* (h^{-1})_{j,l} (1-\delta_{k,l}) \\
    \nonumber
    & \quad + \frac{\left\langle|\xi|^4\right\rangle}{N} \sum_{k} (h^{-1})_{i,k}^* (h^{-1})_{i,k} (h^{-1})_{j,k}^* (h^{-1})_{j,k} \\
    \nonumber
    &= \sum_{k,l} (h^{-1})_{i,k}^* (h^{-1})_{i,k} (h^{-1})_{j,l}^* (h^{-1})_{j,l} \\
    \nonumber
    & \quad + \frac{1}{N} \sum_{k,l} (h^{-1})_{i,k}^* (h^{-1})_{i,l} (h^{-1})_{j,l}^* (h^{-1})_{j,k} + \frac{\left|\left\langle\xi^2\right\rangle\right|^2}{N} \sum_{k,l} (h^{-1})_{i,k}^* (h^{-1})_{i,l} (h^{-1})_{j,k}^* (h^{-1})_{j,l} \\
    & \quad + \frac{1}{N}\left( \left\langle|\xi|^4\right\rangle - 2 - \left|\left\langle\xi^2\right\rangle\right|^2 \right) \sum_{k} (h^{-1})_{i,k}^* (h^{-1})_{i,k} (h^{-1})_{j,k}^* (h^{-1})_{j,k}.
\end{align}
The first term of Eq. (\ref{eq:u_i^2u_j^2}) is equal to $\left\langle \overline{|w_i|^2}\right\rangle \left\langle\overline{|w_j|^2} \right\rangle$. 
The subsequent terms are the covariance. 
The standard deviation of the statistical error in the estimate of the landscape at any particular site decrease as $1/\sqrt{N}$ as the number $N$ of vectors sampled increases.

If the elements of the random vectors are independently and identically distributed with a normal distribution 
with mean zero and variance one, then
\begin{align}
    \left\langle|\xi|^4\right\rangle &= 3,\\
    \left\langle\xi^2\right\rangle &= 1,
\end{align}
and the covariance is
\begin{align}
    \nonumber
    &\mathrm{Cov}\left[\overline{|w_i|^2}, \overline{|w_j|^2}\right] = \left\langle \overline{|w_i|^2}\cdot\overline{|w_j|^2} \right\rangle - \left\langle \overline{|w_i|^2}\right\rangle \left\langle\overline{|w_j|^2} \right\rangle\\
    &\quad = \frac{1}{N} \sum_{k,l} \left[ (h^{-1})_{i,k}^* (h^{-1})_{i,l} (h^{-1})_{j,l}^* (h^{-1})_{j,k} + (h^{-1})_{i,k}^* (h^{-1})_{i,l} (h^{-1})_{j,k}^* (h^{-1})_{j,l} \right].
\end{align}

If the elements of the random vectors have absolute value unity and random phases as in Eq. (\ref{eq:random_phase})
with the phase $\theta_i$ independently and identically distributed with a uniform distribution on $\left[0,2\pi\right]$,
then
\begin{align}
    \left\langle|\xi|^4\right\rangle &= 1,\\
    \left\langle\xi^2\right\rangle &= 0,
\end{align}
and the covariance is
\begin{align}
    \mathrm{Cov}\left[\overline{|w_i|^2}, \overline{|w_j|^2}\right] = \frac{1}{N} \sum_{k\neq l} (h^{-1})_{i,k}^* (h^{-1})_{i,l} (h^{-1})_{j,l}^* (h^{-1})_{j,k}.
\end{align}

\section{} \label{alternative_method}

The $q_k$ appearing in Eq. (\ref{eq:qk}) occur in complex conjugate pairs. Moreover, since the $h_k$ commute an arbitrary reordering of the product in Eq. (\ref{eq:factorization_complex}) is permissible.
We first note that
\begin{equation}
    \left(q+q_k\right) \left(q+q_k^*\right) = q^2 + r_k q + 1,
\end{equation}
where
\begin{equation}
    r_k = q_k + q_k^* = 2\cos\left(\frac{2k-1}{p}\pi\right),
\end{equation}
The product of $2p$ factors in Eq. (\ref{eq:factorization_complex}) can then be expressed using $p$ factors as
\begin{equation}
    h^{2p}+1 = \prod_{k=1}^{p}  h'_k.
\end{equation}
where
\begin{equation}
    h'_k =  h_0\left(h_0 + r_k I\right) + I.
\end{equation}
Now, if $H$ is Hermitian, then so are the $h_k'$.
The conjugate gradient method is now applicable.

\end{document}